# Do you need a blockchain in construction? Use case categories and decision framework for DLT design options

Jens J. Hunhevicz[1]*, Daniel M. Hall[1]

[1]Institute of Construction and Infrastructure Management, Department of Civil, Environmental and Geomatic Engineering, ETH Zurich, Zurich, Switzerland

*Corresponding author

**Abstract**

Blockchain and other forms of distributed ledger technology (DLT) provide an opportunity to integrate digital information, management, and contracts to increase trust and collaboration within the construction industry. DLT enables direct peer-to-peer transactions of value across a distributed network by providing an immutable and transparent record of these transactions. Furthermore, there is potential for business process optimization and automation on the transaction level, through the use of smart contracts, which are code protocols deployed on supported DLT systems. However, DLT research in the construction industry remains at a theoretical level; there have been few implementation case studies to date. One potential reason for this is a knowledge gap between use-case ideas and the DLT technical system implementation. This paper aims to reduce this gap by 1) reviewing and categorizing proposed DLT use cases in construction literature, 2) providing an overview of DLT and its design options, 3) proposing an integrated framework to match DLT design options with desired characteristics of a use case, and 4) analysing the use cases using the new framework. Together, the use case categories and proposed decision framework can guide future implementers toward more connected and structured thinking between the technological properties of DLT and use cases in construction.

## 1. Introduction

*1.1. Distributed Ledger Technology*

The concept of Distributed Ledger Technology (DLT) provides a distributed peer-to-peer system for value transactions without any intermediation from a central authority. The most prominent type of DLT is blockchain, which has its origin in the peer-to-peer cryptocurrency Bitcoin [1]. Bitcoin solved for the first time the double-spending problem through its proof-of-work consensus algorithm [1]. The overarching idea was to timestamp transactions and proof-of-work hashing them into a sequential record (also called *chain*) that cannot be changed without redoing the proof-of-work. As long as the nodes controlling the network and performing the proof-of-work do not collaborate to attack the network, these inherent system properties enable participants to trust that the history of transactions is correct. These properties are called "fundamental properties" of a DLT [2].

Overall, high fundamental properties lead to a more secure and trustworthy system. Of course, this security comes at a cost. There is a tradeoff between performance (in terms of transaction speed and overhead of the system) and the fundamental properties. Therefore, one type of DLT (e.g. Bitcoin) is unlikely to meet the prerequisites for all usage scenarios [2]. Other implementations of DLT have emerged to meet the different implementation requirements. In this paper, DLT *design options* refers to the potential selection of various DLT implementations. Therefore, DLT is an overarching term that captures various potential design options [3,4].

Furthermore, there are other constraints regarding the functionality of certain DLT implementations. Most important, newer DLT implementations enable the use of *smart contracts.* Smart contracts have been popularized by the DLT *Ethereum* [5], which allows the execution of code protocols on the DLT.

Smart contracts enables the automation of business logic for assets and data managed on the DLT. They also enable the creation of new types of "tokenized" digital assets.

To summarize, the fundamental properties of DLT enable the building of trust between transacting parties and devices, as well as the potential to increase the settlement time of transactions and reduction of costs associated with intermediaries [6]. In combination with the functionality of smart contracts, the potential applications of DLT in society and industry are manifold. Industries such as financial services, insurance, and supply chain envision it to be a future game changer on how these sectors interact and transact. Future peer-to-peer interactions and process automation using DLT can be more trustworthy and transparent compared to traditional applications.

Most literature agrees that DLT should not be neglected when looking at future business development (e.g. Tapscott and Tapscott [7] and Nowinski and Kozma [8]). This is a proposition that should also be considered in the construction industry.

*1.2. DLT for the Construction Industry*

While various industries have already developed different DLT prototypes and applications, the construction sector is only at the beginning of DLT implementation as a tool. However, the application of DLT in construction might be especially promising [9,10]. In contrast to many other industries, the construction industry structure can be characterized as a decentralized, loosely-coupled network. This leads to various unique challenges regarding its structure [11]. Construction is delivered by project teams that work in cross-functional, geographically distributed teams [12] composed of complex and fragmented supply chains [13]. The successful completion of complex projects requires the development of trust and mutual confidence between the interacting parties for each individual project [14]. This has been found to be a major challenge for large, complex, and long-term projects that rely on the interdependent actions of numerous stakeholders [12,15]. Mistrust leads to guarded behaviors and conflicts within project teams. It often results in individuals pursuing and protecting their own interests instead of the benefit of the overall projects [14]. Furthermore, without a strong foundation of trust, it is difficult to reach consensus and information exchange in a meaningful manner [16].

To summarize, the decentralized and project-based structure of the construction industry requires many stakeholders with various incentives to interact over long time horizons. This leads to coordination challenges such as a lack of trust, poor information exchange, and supply chain fragmentation. In theory, the potential benefits of DLT to provide a trusted means for transactions aligns with these coordination challenges. DLT can help by making construction more efficient, transparent, and accountable between all involved participants [10]. However, despite theoretical alignment of DLT value propositions and coordination challenges in construction, there are few implementations of DLT in a construction context.

Most literature to date instead provides an overview of the potential *use cases* for DLT in construction. For example, early literature sees the vision for DLT as a complementary technology to building information modelling (BIM) and internet of things (IoT) [10,17–19]. BIM allows designers and builders to design, visualize, and coordinate construction systems with greater efficiency through the use of three-dimensional modelling tools and processes. While helpful for individual firms, BIM provides significantly more value when it can integrate information across multiple firms and organizations in the supply chain [20]. Despite its potential, the adoption of BIM has lagged as project teams struggle with trust and liability concerns associated with sharing information on the project [21–23]. It seems that new technologies such as BIM that promise to increase collaboration in the construction industry are again hindered by issues of trust and liability found throughout the industry [21,24]. IoT describes an environment where physical objects connect with the digital world using sensors and connected devices [25]. Ye et al. [19] see DLT as a way to hold the data produced by IoT in a transparent, secure and convenient environment and BIM as the baseline tool to digitize the construction project data. De La Pena and Papadonikolaki [26] suggest that the combination of DLT and IoT can increase inter-firm trust in construction. Eventually, this could lead to a future industry state characterized by the "circular economy of BIM things" [10,18]. The produced data from projects and IoT can be integrated into a common data environment – first developed and visualized through BIM during design and construction – enabling a digital twin consistently maintained over the whole life cycle of a building.

DLT acts as an immutable track-record for higher transparency and potential automation through smart contracts.

*1.3. Goal and Scope of the Study*

The mentioned vision for DLT use cases is ahead of the current state of research, since very few documented implementations of DLT for the construction industry exist. There is now need for prototypes and use-case implementations to assess and validate these value propositions for DLT in construction. More specific use cases on how DLT can be used in construction have been proposed by various authors. Some of them can align with the above vision and rely on combination with BIM and IoT, but some can also stand on their own. Little research has attempted to structure these use cases into categories according to the different value propositions of DLT. A categorization for use cases might help to more easily align the prerequisites of specific use cases with the needed DLT design options, since use cases in construction have been mostly understood at the theoretical level and often lack a detailed understanding of the technical system implementation [19]. Most importantly, the DLT design option with its fundamental properties should match the trust requirements of the proposed use case. On top of that, other constraints regarding technical capabilities of the needed DLT should be considered. Finally, the fast-moving and vast landscape of DLT is challenging for potential implementation of the diverse DLT use cases in construction. There is need of a framework so that researchers looking to implement DLT for a use case can start by choosing an appropriate system. Without a good understanding of both use case function and DLT design options, it can be difficult for implementers to begin development of a proof-of-concept for a use case.

This paper aims to reduce the gap between DLT use cases and DLT technical system implementation in construction. To do so, the paper first reviews and categorizes DLT use cases proposed in existing literature into higher level categories aligned with the specific value propositions of DLT. Second, the paper describes the technical features of DLT and from this summarizes four different DLT design options next to traditional database solutions. Third, the paper proposes a decision framework to answer the question "do you need a blockchain in construction?" and if so, which type of DLT design option should be selected. Fourth, the paper uses the framework to evaluate each proposed use case and reports the potential DLT design options that could be used. Finally, interesting findings are discussed and limitations stated.

## 2. Categorization of DLT Use Cases in Construction

A number of papers and consultancy reports started in 2017 to identify potential use case scenarios to deploy DLT in the construction sector. A review of fifteen sources (see Table 1) identifies the potential use cases proposed for DLT in construction. Because literature on DLT in construction is still limited, both scholarship and consulting reports are considered. The review scope is limited to literature focusing on the construction industry and excludes literature about the energy sector, smart cities and homes, and very general work about the built environment.

This review of the DLT use case literature identifies twenty-four potential use cases. These cases can be further clustered into higher-level use case categories (see Table 2). Table 2 provides a summary of the categorized use-cases by source. This is an extension and update of the use case categorization originally performed by Hunhevicz and Hall [27] with addition of six relevant recently-published papers. Furthermore, a specific refinement of Hunhevicz and Hall [27] is made by splitting the use case category of "record of transactions, changes, ownership" into two separate categories related to "immutable records of transactions" and "immutable records of assets/ownership" (Table 2, Categories 3 & 4).

On a high level, the categories shown in Table 2 are in line with the main value propositions of DLT:

1) Higher transparency and trust in the project and supply chain due to the fundamental properties of DLT (Table 2, category 3, 4).
2) Higher efficiency and accuracy in business process optimization and automation through the use of smart contracts (Table 2, category 1, 2, 6, 7), as well as the use of smart contracts to create tokens for financial, incentive, or other purposes (Table 2, category 5).

*Table 1: Literature for use-case analysis (S: scholarly papers, C: consulting reports)*

| # | Author (Year) | Title | Type |
|---|---|---|---|
| [1] | Belle [28] | The architecture, engineering and construction industry and blockchain technology | S |
| [2] | Heiskanen [29] | The technology of trust: How the Internet of Things and blockchain could usher in a new era of construction productivity | S |
| [3] | Kifokeris and Koch [30] | Blockchain in construction logistics: state-of-art, constructability, and the advent of a new digital business model in Sweden | S |
| [4] | Kinnaird and Geipel [18] | Blockchain Technology: How the Inventions Behind Bitcoin are Enabling a Network of Trust for the Built Environment | C |
| [5] | Li et al. [9] | Blockchain in the built environment and construction industry: A systematic review, conceptual models and practical use cases | S |
| [6] | Li et al. [31] | A Proposed Approach Integrating DLT, BIM, IoT and Smart Contracts: Demonstration Using a Simulated Installation Task | S |
| [7] | Luo et al. [32] | Construction Payment Automation through Smart Contract-based Blockchain Framework | S |
| [8] | Mason [33] | Intelligent Contracts and the Construction Industry | S |
| [9] | Mathews et al. [17] | BIM+Blockchain: A Solution to the Trust Problem in Collaboration? | S |
| [10] | Nawari and Ravindran [34] | Blockchain and the built environment: Potentials and limitations | S |
| [11] | O'Reilly and Mathews [35] | Incentivising Multidisciplinary Teams with New Methods of Procurement using BIM + Blockchain | S |
| [12] | Penzes [10] | Blockchain technology: could it revolutionise construction? | C |
| [13] | Turk and Klinc [36] | Potentials of Blockchain Technology for Construction Management | S |
| [14] | Wang et al. [37] | The outlook of blockchain technology for construction engineering management | S |
| [15] | Ye et al. [19] | Cup-of-Water theory : A review on the interaction of BIM, IoT and blockchain during the whole building lifecycle | S |

*Table 2: Use-Case clustering into seven categories, based on the literature listed in Table 1 (adapted from Hunhevicz and Hall [27]).*

|     | Use-Case Category | Literature | | | | | | | | | | | | | | |
| --- | --- | --- | --- | --- | --- | --- | --- | --- | --- | --- | --- | --- | --- | --- | --- | --- |
|     |                   | [1] | [2] | [3] | [4] | [5] | [6] | [7] | [8] | [9] | [10] | [11] | [12] | [13] | [14] | [15] |
| 1   | Internal Use for Administrative Processes | | | | | | | | | | | | | | | |
| 1.1 | *Notarization and Synchronization of Documents* | | | | | | | | | | | | | | X | |
| 2   | Transaction Automation between Stakeholders with Smart Contracts | | | | | | | | | | | | | | | |
| 2.1 | *Triggering Payments* | | | X | X | X | X | X | X | | | | X | | X | X |
| 2.2 | *Triggering Contract Deliverables* | | | | X | | | | X | | X | | X | | | |
| 2.3 | *Self-executing Contract Administration* | | | | | | | X | | | | | X | | X | |
| 2.4 | *Automated Data/Information Sharing* | | | | X | | | | | | | | X | | | X |
| 2.5 | *Automated Code Compliance Checking* | | | | | | | | | | X | | | | | |
| 3   | Immutable Record of Transactions | | | | | | | | | | | | | | | |
| 3.1 | *Timestamping of "Value" Transactions* | | X | | | X | | | | X | X | X | | | | |
| 3.2 | *Record of Changes in digital models (BIM)* | | | X | X | X | | | X | | X | X | X | X | | X |
| 3.3 | *Tracking of Supply Chain Logistics* | | X | X | X | | | | | | | | X | | X | X |
| 3.4 | *Tracking of Project Progress and Worked Hours* | | | | | | | | | | | | X | | | |
| 3.5 | *Record of Maintenance and Operations Data* | | | | | | | | | | | | X | | | |
| 3.6 | *Tracking of Health & Safety Incidents* | | | | | | | | | | | | X | | | |
| 3.7 | *Verification of Installation Tasks* | | | | | | X | | | | | | | | | |
| 3.8 | *Record/Notarization for Regulation and Compliance* | | | | | X | | | | | | | | | X | |
| 4   | Immutable Record of Assets/Ownership | | | | | | | | | | | | | | | |
| 4.1 | *Record of Ownership in BIM (IP-Rights)* | X | | | X | X | | | X | | X | | X | X | | X |
| 4.2 | *Record of Ownership for Physical Assets (e.g. Property)* | | | | X | | | | X | | | | X | | | X |
| 4.3 | *Managing Identities for Reputation (People, Contractors)* | X | | | X | | | | | | | | X | | | |
| 4.4 | *Material & Product Passports (Provenance and Properties)* | | | | X | | | | | | | | X | | | |
| 5   | Coins/Tokens as Payment or Incentive Scheme | | | | | | | | | | | | | | | |
| 5.1 | *Payment in Cryptocurrencies* | | | | | | | | | | | | | | X | X |
| 5.2 | *Shared Accounts & Insurances* | | | | | X | | | X | | | | | | | |
| 5.3 | *Incentives over the Whole Building-Lifecycle* | | | | | | | | | | X | | X | | | |
| 6   | Decentralized Applications (DApps) | | | | | | | | | | | | | | | |
| 6.1 | *Decentralized Market Places for Products and Services* | | | | X | | | | | | | | | | X | |
| 6.2 | *Decentralized Common Data Environments (CDE) for Digital Models* | | | | X | | | | | | X | | | | | X |
| 7   | Decentralized Autonomous Organizations (DAOs) | | | | | | | | | | | | | | | |
| 7.1 | *Automated Building Maintenance Systems* | X | | | | | | | | | | | X | | 7.1 | X |

*1 - Internal Use for Administrative Purposes*

DLT can be used for **notarization and synchronization of documents** (Table 2, 1.1). This includes the storage and perfect notarization of each creation, deletion, and updating of files across an inter-organizational system [37]. This can simplify and automate administrative processes. Wang et. Al. [37] mentions the recording of quality data or resource consumption data as examples.

*2 - Transaction Automation Between Stakeholders with Smart Contract*

Using smart contracts, DLT can automate transactions between different stakeholders. The most mentioned use case is automatic **triggering payments** (Table 2, 2.1). This is helpful because delays for monetary transactions are mentioned repeatedly as a factor causing conflicts and disputes (Eastman 2011). In addition, automatic **triggering contract deliverables** are mentioned multiple times, where an updated state in the ledger causes a predefined contractual action (Table 2, 2.2). Once a smart contract is written, its behavior is unambiguous and predictable. This can be used for **self-executing contract administration** (Table 2, 2.3), such as monitoring and updating of the contract status [37]. Smart contracts are also mentioned as a way to enable **automated information and data sharing** in projects (Table 2, 2.4), ensuring consistent reporting for (sub)contractors and owners. Finally, Nawari and Ravindran [34] introduce a framework for **automated code compliance checking** (Table 2, 2.4) in the BIM design review process. All use cases are independent of the construction project phase and can be applied for procurement and supply chain activities for higher accuracy and efficiency.

*3 - Immutable Record of Transactions*

DLT can provide immutability and transparency for transactions. On a high level, DLT can provide a **timestamping of value transactions** (Table 2, 3.1). The most mentioned use case is the **record of changes in digital models**, especially in combination with BIM (Table 2, 3.2). One other often mentioned use case is the **tracking of supply chain logistics**, including procurement, transportation, and storage of goods (Table 2, 3.3). Penzes [10] expands on the tracking of processes towards **tracking of project progress and worked hours** (Table 2, 3.4), **maintenance and operations data** of buildings and machines (Table 2, 3.5), and **health & safety incidents** (Table 2, 3.6). Li et. al. [31] describes **verification of installation tasks** as a use case for DLT, in particular correct installation of insulation panels (Table 3, 3.7). Finally, two papers [9,37] describe the **record/notarization for regulation and compliance** as potentially advantageous in construction (Table 2, 3.7).

*4 - Immutable Record of Assets/Identities*

As in the use case category 3, the focus lies on the immutability and transparency provided by DLT. In addition to recording transactions, DLT can also record information of physical or digital assets. One potential use case mentioned is the **record of ownership in BIM** for IP-protection (Table 2, 4.1). If not a digital asset, a unique digital counterpart of the respective physical asset can be created. For example, a **record of ownership for physical assets** such as property (Table 2, 4.2). Furthermore, **managing identities for reputation** of people or organizations on DLT (Table 2, 4.3) for clear and trustworthy identification is possible. Similarly, **material and product passports** with product and provenance-related information (Table 2, 4.4) can be maintained throughout the supply chain. This can be used for quality assurance in global construction projects [37] or to enable the reuse of materials at a later stage of a building towards a circular economy [18]. Also, certification of products and buildings could profit from the availability of this trusted data.

*5 - Coins/Tokens as Payment or Incentive Scheme*

DLT enables new financial and incentive related use cases by creating coins or tokens. A well-documented use case is **payment in cryptocurrencies** (Table 2, 5.1). This allows participants to send money across borders instantly and with small transaction fees. This can be extended even further with shared risk and reward structures for **shared accounts and insurances** among multiple, independent stakeholders (Table 2, 5.2). Finally, Mathews et al. [17] propose the use of an #AECoin as a token to provide **incentives over the whole building life-cycle** to reward project contributors for the contributed value even after project handover to the client (Table 2, 5.3). This can create superior value for the project owner, as participants can be incentivized to make long-term life-cycle decisions in order to increase their own rewards. Similarly, O'Reilly and Mathews [35] describe a DLT based incentive approach in BIM in order to create more energy efficient buildings and save energy in the use phase.

*6 - Decentralized Applications (DApps)*

DApps are applications that are based on a DLT that is not run by any intermediary. This means that no censorship of users beyond rules encoded in the smart contracts is possible. DApps enable direct user interaction with DLT, typically through web user interfaces. Even though it is possible to create web applications for the use cases in the previous categories for very project-specific cases, this category refers to DApps for long-term and global users across project boundaries. Users of such applications might be unknown and involved in various projects simultaneously. Different use cases for DApps are mentioned in the literature. **Decentralized marketplaces for products and services** (Table 2, 6.1) can be set up based on digital identities (Table 2, 4.3). This can enable access to objective data (e.g. the most-qualified person or company in tendering) without the need to disclose sensitive data to third parties [28]. Also, **decentralized common data environments (CDE) for digital models** as a combination of cloud storage and DLT are proposed to store digital models without the need to trust a third party server provider or run private servers vulnerable to attacks [19].

*7 - Decentralized Autonomous Organizations (DAOs)*

DAOs represent a fully autonomous organization based on smart contracts that run on DLT without any human involvement. Governance rules are coded in smart contracts and incentive mechanisms are implied through crypto-economic design (CED). Often, DAOs make use of IoT to interact with the real world and a digital model to provide location context. Even though fully automated construction companies seem futuristic, three sources [10,19,28] mentioned **automated building maintenance systems** as one possibility for a DAO (Table 2, 5.1). The idea is that building performance can be monitored through sensors (IoT) in combination with BIM. This enables an automatized reaction to certain conditions based on predefined rules. Specific examples include the automatic ordering of spare parts or regulating technical installations based on predefined performance indicators.

## 3. Overview of DLT Technology and Design Options

After having categorized use cases in construction into higher level categories aligned with different value propositions of DLT, technical aspects of DLT and how they relate to the fundamental properties need to be introduced in order to design a connecting framework. This helps to understand the relationship of technical DLT-features with the different expectations of use cases regarding their capabilities.

### 3.1. DLT Technology Stack

While a full explanation of the underlying base technologies of DLT is beyond the scope of this paper, this section provides an overview of the most important factors that influence DLT design options. This section sources from more detailed explanations of DLT (e.g. Wattenhofer [38]) and scholarship that introduces taxonomies for DLT while providing in-depth explanations on different components [2,39,40]. Information is structured based on an adapted version of the technology stack used in Shermin [41], pictured in Figure 1. The *internet layer* acts as the base technology for information sharing. A DLT, sometimes also referred to as *protocol layer* [3], is built on top of *the internet layer* with three main components impacting its characteristics: Ledger, Peer to Peer (P2P) Network, and Governance. If code can be executed on the protocol layer, an *application layer* is possible with smart contracts.

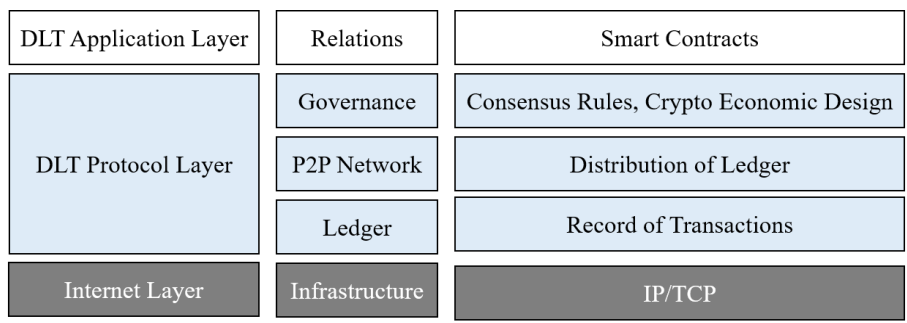

*Figure 1: Technology stack of DLT (adapted from Shermin [41])*

*3.1.1. Ledger*
The ledger represents the data structure of DLT. The most-well known ledger type as in the case of Bitcoin is a *blockchain* with sequential entries and total order [39]. Blockchain links the latest block containing the most recent transaction information with the previous blocks to create a "chain". Integrity of the ledger is reached through the process of hashing, applying mathematical one-way functions repeatedly to the transaction data. These hashes are included in a block together with the block-hash of the previous block, making it possible to notice if past data has been tampered. With every new block, the chances of attacking a previous block decrease exponentially [1]. Besides blockchain, other types of ledgers are possible. For example, the *directed acyclic graph* (DAG) is a ledger with a stream of individual transactions entangled together that can be confirmed in parallel (e.g. IOTA [42]). Typically, there is only one ledger per DLT. However, new research focuses on how to process transactions in more than one ledger (*sidechains* as e.g. in Back et al. [43]) or among multiple smaller groups of nodes (*sharding* as e.g. in Zamani et al. [44]) in a network to make it more scalable.

Various elements of a ledger can be defined such as the storage capabilities or data encryption. Next to the defined size of a block or transaction, the ledger can store the default transaction information and/or additional data. Transactions on the ledger are usually encrypted through hashing, but might be still *linkable* and therefore reveal further information about the sender and receiver. Some systems allow *obfuscatable* transactions by using advanced cryptography (for an overview and comparison of existing systems see e.g. Yocom-Piatt [45]). Encrypted transactions and data become important for privacy considerations in *public* DLT systems (see section 3.1.2).

Finally, if the ledger supports turing complete language on the protocol layer, an application layer for coded relations is possible (see Figure 1). This enables the use of *smart contracts*, described the first time by Szabo [46]. Smart contracts represent code protocols that execute certain logic based on the state of the ledger. The name "smart contracts" can be misleading. They do not represent a contract per se, but could be coded in such a way. Since they run on a DLT, the code is also unchangeable unless programmed to be updateable. These smart contracts can be used to create autonomous work flows or containers of value (e.g. representing currencies, securities, utilities, or other), so-called *tokens* [47]. Many smart contracts can be combined to build so-called decentralized applications (DApps) or decentralized autonomous organizations (DAOs) (see also Section 2).

*3.1.2. P2P Network*
The ledgers are distributed on different nodes in the network. Setting up these nodes can be either permissionless or permissioned. *Permissionless nodes* allow anyone to set up a node and write transactions to the ledger by participating in the consensus mechanism (see section 3.1.3). *Permissioned nodes* cannot be set up by anyone and/or limit write-access to the ledger. The second distinction is between *public* and private ledgers in the network. *Public ledgers* allow anyone to read the ledger. *Private ledgers* allow only defined members to access transactions on the ledger. The distribution and ownership of nodes impacts the decentralization of the system. *Public permissionless* DLT naturally lead to higher network decentralization. Because anyone can set up a node, this leads to more nodes and a higher variability in the interests of the participating users. Typically, data is replicated on all participating nodes. However, there exists DLT design options that do not replicate data on all nodes but only on nodes that are allowed to access the data (e.g. in Corda [48] or Holochain [49]).

*3.1.3. Governance*
The governance of the DLT defines the set of rules for users interacting with the system. The most important component is the consensus mechanism. The *consensus mechanism* is responsible for defining how to write, validate, and agree on entries to the ledger. Proof-of-work was the first blockchain consensus mechanism and the greatest innovation behind Bitcoin (see Nakamoto [1]), protecting the network effectively from double-spending and attacks to ensure immutability and non-repudiation of data [50]. In the case of proof-of-work, the honest nodes need to control the majority (> 50%) of CPU power to protect the network. The more network decentralization, the less likely it becomes that nodes can collaborate to attack the network. Since proof-of-work is very resource intensive, other types of consensus mechanisms have been introduced such as proof-of-stake, where nodes validating and adding transactions need to put money at stake that they can lose if they behave dishonestly (see e.g. Tasca and Tessone 2017). All types of consensus mechanisms in *public* DLT are enabled by a crypto-economic design (CED) [51]. A native coin of the DLT incentivizes participants to behave in the interest of the system (e.g. *bitcoin* in Bitcoin or *ether* in Ethereum). This is important to prevent attacks, but also to compensate nodes that validate and add transaction (sometimes called *miners*) for their expenditures. A successful CED incentivizes honest behaviour in a DLT network. Multiple properties of a CED can be defined, influencing the DLT's governance (see also Ballandies et al. 2018). A *private* DLT might not necessarily need a CED, as consensus is often based on permissions (e.g. practical byzantine fault tolerance by Castro and Liskov [52]). This can have an impact on the cost structure for users when interacting with different systems. Often, users pay for transactions on a *public* DLT with transaction fees in its native token. In contrast, users do not have to pay for transactions on a *private* DLT. Costs are predominantly accrued in the acquisition and maintenance of the infrastructure, while making transactions involves usually no fee.

*3.2. Fundamental Properties*

The reason why a DLT is used is given by its fundamental properties. Fundamental properties of DLT are immutability, non-repudiation, integrity, transparency, and equal rights [2]. If the network is decentralized and protected through a working consensus-mechanism, the ledger is *immutable*. Each transaction is added only once to the ledger, which leads to *non-repudiation* of the stored data. The cryptographic tools used on the ledger support data *integrity*, allowing to verify that all the data is complete and as initially written into the ledger. Public access of ledgers for everyone ensures *transparency*, and *equal rights* allow every user the same ability to read and write to the ledger. Table 3 gives a summary of the five fundamental properties.

*Table 3: Fundamental Properties of DLT.*

| Fundamental Property | Explanation |
| --- | --- |
| Immutability | The ledger cannot be tampered after transactions were added. |
| Non-repudiation | Each transaction is added only once to the ledger. |
| Integrity | Data can be verified to be complete and as initially written to the ledger. |
| Transparency | Transactions and data are visible to everyone. |
| Equal Rights | Everyone has the possibility to read and write transactions. |

Trust in the DLT is achieved because the participants rely on the fundamental properties of a DLT itself rather than on trusted third-parties. Different DLT design options exist with varying fundamental properties. Table 4 (inspired by Xu et. al. [2]) summarizes this for central databases and four typical design options of DLT: *private permissioned, private permissionless*, *public permissioned*, and *public permissionless*. The more permissions, the less trust in the technical system can be accomplished with lower overall fundamental properties. This missing trust in the system needs to be compensated by more trust in the participating users or a third party. In some use cases, this high trust in the technical system might not be needed. A more centralized system offers a better performance, as fewer nodes and/or less resource intensive consensus algorithms are used. In addition, privacy can be of concern with public DLT. For example, on-chain data encryption can have insufficient protection, encryption might not be appropriate for a use case, or parties might want to have the possibility to control more aspects of the DLT on the protocol layer (e.g. for easier implementation of system changes).

The relationship of the five different design options can be related to the five fundamental properties (see Table 4). The only fundamental property unaffected by permissions is *integrity* of the data because it is ensured through the cryptographic hash-functions used in all DLT design options. All aspects of a *centralized* database are controlled by a third party, and hence they do not meet any of the fundamental properties. In contrast, *public permissionless* DLT is able to achieve the highest level of trust by maintaining all five fundamental properties. *Public permissioned* DLT restrict write access or even the set-up of nodes and hence do not maintain *equal rights* for all users. In addition, *private permissioned* DLT further limit read access of the ledger and are therefore not *transparent* to users outside the network and inside the network without read-permissions. Furthermore, these permissions might have an impact on the *immutability* and *non-repudiation* of data, since depending on the set up of the DLT governance, outsiders have no assurance when shown the ledger that it has never been modified by the majority of network users (this is why a conditional "yes" (y) was used in Table 4). However, this might be irrelevant to network participants that trust their DLT governance and/or the participating users. Finally, there is the emerging case of *private permissionless* DLT design option not considered by Xu et. Al. [2], where private records can be pegged to permissionless ledgers for proof-of existence [53]. For example, Holochain [49] uses private ledgers connected through distributed hash tables (DHT) [54] to validate data. With this, nodes can be set up in a permissionless way and start interacting with other nodes by only sharing defined information of the private ledger. The DHT ensures *non-repudiation* and *immutability* of the shared data (but not the private data). Furthermore, *equal rights* are guaranteed since the network is permissionless. But since read access is limited to shared data, *transparency* to anyone is not ensured.

*Table 4: The inversely related impact of the fundamental properties and performance in different design option (n: no; y: yes).*

| Design Option | Comment | Examples | Impact - Fundamental Properties | | | | | Overall | Performance |
|---|---|---|---|---|---|---|---|---|---|
| | | | Immutability | Non-repudiation | Integrity | Transparency | Equal Rights | | |
| Centralized | Central databases with a single or alternative providers | - | n | n | n | n | n | ↓ | ↑ |
| Private Permissioned DLT | DLT with permissions on both read & write-access | Hyperledger Fabric[1], Corda[1] | (y) | (y) | y | n | n | | |
| Private Permissionless DLT | DLT with permissioned read-access & permissonless write-access | Holochain[2] | y | y | y | n | y | | |
| Public Permissioned DLT | DLT with permissionless read-access & permissions for write-access | EOS[1] | Y | y | y | y | n | | |
| Public Permissionless DLT | DLT with permissionless read access & permissionless write-access | Bitcoin[1], Ethereum[1] | y | y | y | y | y | | |

[1] *Examples classified by Ballandies et. al. (2018): Ethereum (www.ethereum.org), EOS (www.eos.io), Hyperledger Fabric (www.hyperledger.org/projects/fabric), and Corda (*www.r3.com*).*
[2] *Example classified by Daniels [55]: Holochain (www.holochain.org).*

## 4. A Decision Framework for DLT Design Options in Construction

### 4.1. Review of Existing Frameworks

Decision frameworks for DLT aim to guide users to the best-suited DLT design option for their use case in a structured way. Overall, many factors can be considered with a large solution space. This is aggravated by the fact that the technical landscape of DLT is fast moving and changing. However, some contributions already dealt with this question. Eight sources were identified (Table 5) and analyzed regarding their approach.

*Table 5: DLT decision frameworks and their different approaches to determine the right DLT design option.*

| | Source (↑ Publication Date) | Type | Inputs | Outputs |
|---|---|---|---|---|
| [a] | Peck [56] | Sequential Framework | Seven questions related to: *Participants, Likelihood of Attack, Trust, Possibility of Third Party, Privacy, Updateability of Data.* | Three options: *No DLT, permissioned DLT, public DLT.* |
| [b] | Turk and Klinc [36], based on Suichies [57] | Sequential Framework | Eight questions related to: *Possibility for Traditional Database, Trust, Alignment of Interests, Possibility of Third Party, Control of Functionality & Privacy, Type of Consensus.* | Four options: *No DLT, public DLT, hybrid DLT, private DLT.* |
| [c] | Xu et al. [2] | Sequential Framework | **Trusted** authority, Ability to Decentralize Authority, Various Technical Configurations, Other Design Decisions | DLT, Traditional Database |
| [d] | Rangaswami et al. [58] | Sequential Framework | 11 questions related to: *Possibility of Traditional Database, Technical Limitations, Relationship of Participants, Trust, Control of* | Five options: *No DLT, not ready for DLT applications, further research needed, private DLT, public DLT.* |
| | | | *Functionality*. | |
| [e] | Wessling et al. [59] | Four Steps | Step 1: Identify participants.<br>Step 2: **Trust** relations.<br>Step 3: Interactions. | Step 4: Derive system architecture by overlaying trust and interactions. |
| [f] | Wüst and Gervais [60] | Sequential Framework | Six questions related to: *Database Type, Participants Known & **Trusted**, Alignment of Interests, Need for Public Verifiability*. | Four options: *No DLT, private permissioned DLT, public permissioned DLT, permissionless DLT*. |
| [g] | Hunhevicz and Hall [27] | Mapping Based on **Trust** Proxy | Three questions to determine the proxy "*level of **trust***" in a use case. Table with fundamental properties of the DLT design options. | Four options: *Fully centralized, private DLT, public permissioned DLT, public permissionless DLT*. |
| [h] | Li et al. [9] | Sequential Framework | 14 questions: a combination of Peck [56] and Rangaswami et al. [58]. | Five options as in Rangaswami et al. [58]. |

*4.2. Proposed Stages for Construction Decision Framework*

An integrated framework was created pictured in Figure 2, combining the analyzed approaches (Table 5). The most frequent connection between the analyzed framework was the consideration of trust as a criteria to decide on a DLT. Since relying on trust as the main decision criterion, the authors base the main idea of the framework on the approach of Hunhevicz and Hall [27]. An assessment of the trust relations in a use cases is made according to the fundamental properties needed by the DLT design option. This leads to an optimization of the chosen solution regarding the performance of the system, while ensuring that the chosen DLT option actually provides the needed properties. Wessling et al. [59] also follow this procedure; participants and interactions are determined first and then the network architecture is designed. For the more detailed structure of the framework the approach of Wüst and Gervais [60] is used for two reasons. First, the framework is aligned with the chosen approach to assess first the fundamental properties needed for a use case (Stage 1 – Do you need a DLT?). Second, it is the most extensive in terms of outputs of DLT design options (Stage 2 – which DLT design option?). Each question or evaluation step of the other frameworks were cross-compared and the framework of Wüst and Gervais [60] was modified where chosen appropriate. Modifications include the addition of question 4, renaming question 7 & 8, and adding a third stage to consider other important, mostly technical constraints (see Figure 2). To be complete with the introduced DLT design options (see Table 4), the private permissionless DLT option was added by the authors including question 7 and 9 (see Figure 2). This was not considered by any other framework reviewed (Table 5). The detailed reasoning and sources are given in the explanations below.

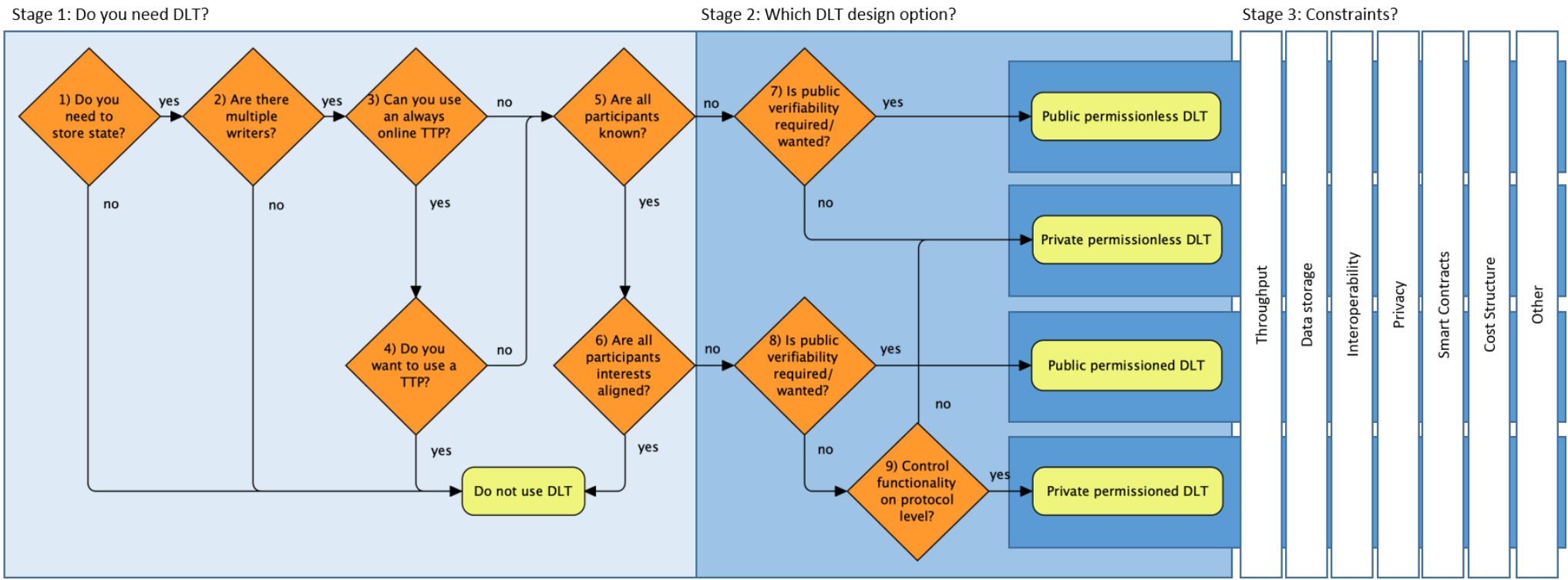

*Figure 2: Framework to decide for a DTL design option based on their fundamental properties in three stages (TTP = trusted third party).*

*4.2.1. Stage 1: Do you need DLT?*

The first stage intends to evaluate whether DLT is needed or no/another database is better suited. It is based on the framework from Wüst and Gervais [60] with slight modifications, using its three more fine-grained questions instead of just one general question whether another database can be used (as proposed by the frameworks in Table 5, [d], [h]). In addition, question 4 was added from the frameworks of Rangaswami et al. [58] and Li et al. [9], and is in line with the question from Xu et al. [2] whether a trusted authority can be decentralized.

1) *"Do you need to store state?"* If storing state is not a requirement, a database is not needed (Sources: Table 5 [b], [f]).

2) *"Are there multiple writers?"* Without multiple writers requiring shared write access, a regular database provides better performance (Sources: Table 5 [d], [f], [h]).

3) *"Can you use an always online trusted third party (TTP)?"* A TTP could operate as a verifier for state transactions using traditional databases with better performance. (Sources: Table 5 [f]).

4) *"Do you want to use a TTP?"* Other reasons such as avoiding intermediaries might be more important than better performance. (Sources: Table 5 [c], [d], [h]).

After question 4), the relationship of the involved participants and their trust setup needs to be assessed with question 5) and 6), asking whether participants are known and whether their interests are aligned. If these two question can be answered with "yes", DLT is not needed. Some questions that appeared in the analyzed frameworks were not considered, since it is already covered by one of the above questions or it is not a finite criteria to use DLT. These are: *"Use case deals with digital assets?"* (Table 5 [d], [h]), *"Permanent record wanted?"* (Table 5 [d], [h]), *"Manages contractual or value exchange?"* (Table 5 [d], [h]).

*4.2.2. Stage 2: What DLT design?*

Stage 2 of the framework evaluates the best suited DLT design option for a use case. Notably, all analyzed sources (Table 5) mention the trust setup of the participants to decide for a certain DLT design option. As discussed previously (chapter 3.2), DLT can be seen as a mean to manage missing trust relations in a use case through the implied fundamental properties (Table 4). The reviewed frameworks vary in their approach to trust. Peck [56] and Wüst and Gervais [60] (Table 5 [a], [f]) only ask whether the parties are trusted and leave to the reader what trust means. Rangaswami et al. [58], Hunhevicz and Hall [27] and Li et al. [9] (Table 5 [d], [g], [h]) split it into two questions asking whether contributors are known (which is a separate question also in Wüst and Geravais [60]), and if there interests are aligned. Wüst and Gervais [60] further links the two questions to different DLT design options by relating them to write and read operations on the DLT. Finally, the approach of Wüst and Gervais [60] was used with slightly reformulated questions taken from the other frameworks. First, it is investigated whether a permissioned or a permissionless system is better suited:

5) *"Are all participants known?"* If not, a *permissionless* DLT is suited, since the system allows everyone to join the network and write transactions. (Sources: Table 5 [f]).

6) *"Are all participants interest aligned?"* If the participants are known, but interests are not aligned, a *permissioned* system offers better performance. (Sources: Table 5 [f])

Next, whether a public or private DLT is better suited:

7)/ 8) *"Is public verifiability required/wanted?"* *Public* DLT allow everyone to see transactions in the ledger, *private* DLT have permissions on the visibility and accessibility of data. (Sources: Table 5 [f]).

Since data can be kept private in both *private permissioned* and *private permissionless* DLT, the main difference is the added control on the protocol level in the first. *Private permissioned* DLT need to be run as an own network with all necessary infrastructure. If this is not needed, using a *private permissionless* DLT could be considered, since the network already exists and can be joined setting up a node. Therefore, a question whether participants need to control functionality on the protocol level was included as proposed by some of the frameworks:

9) "Control functionality on protocol level?" *Private permissionless* could be an alternative to *private permissioned* networks if control on protocol level is not needed. (Sources: Table 5 [b], [d], [h]).

*4.2.3. Stage 3: Constraints?*

The framework in stage 1 and 2 is based on the assumption that a DLT design option should be chosen based on the needed fundamental properties in a use case, which are in general inversely related to the performance of a DLT (see Table 4). This approach assumes that performance should be optimized. It means that a better performing DLT will be chosen, if the higher fundamental properties of DLT are not required. This decision is in the end directly related to the security of the system. More decentralized, public systems protected by strong consensus mechanisms allow for high security of data without the need to trust an intermediary (see section 3.2). Choosing more permissioned systems might bring other benefits (e.g. higher throughput), but compromise the fundamental properties of the system (less security).

Having said that, the decision might shift to another DLT design option, if more importance on other factors is placed. Therefore, stage 3 is introduced in the framework to assess other constraints. For example, the frameworks of Rangaswami et al. [58] and Li et. al [9] (Sources: Table 5 [d], [h]) have limited throughput and storage of large amounts of non-transactional data as a question at the beginning, excluding use cases from using DLT if this holds true. In contrast, the proposed framework (Figure 2) analyses first if DLT is suited for the use case and investigates then in stage 3 whether there are constraints that are problematic for a use case. This is proposed because of the following reasons:

- Constraints, especially technological ones, are subjective to fast progress and change. A framework including them early in the evaluation is likely to be outdated soon.
- The proposed constraints in stage 3 can be adapted based on the use case, leading to a flexible framework.
- There is an emerging ecosystem around DLT, where DLT is seen as only part of the bigger technology stack. This will increase the possible solution space, where some limitations of DLT can be solved through alternative technologies interacting with it.

In Table 6, six constraint dimensions that could be considered for a final DLT solution are proposed. They are partially based on the technical considerations in the framework of Xu et al. [2] and other reviewed literature and do not claim to be complete. Hence, a dimension "Other" to account for any constraint relevant to a use case not captured by the six dimensions is included. Often, to have all benefits in one system is not possible and compromises need to be made based on the use case requirements.

*Table 6: Proposed constraint dimensions for stage 3 in the framework.*

| | |
|---|---|
| Throughput: | Throughput is an important constraint for DLT applications and is known to be a limitation for certain DLT design options. Throughput is generally contradicting decentralization of DLT. More centralized systems offer better performance. Next to variations on the protocol layer (such as the data structure, ledger type, and consensus protocols), possible solutions are sharding or side-chains (see section 3.1.1). If off-chain transaction are anchored to an existing DLT, they are referred to as 2nd layer solutions. Examples are the plasma side-chain for Ethereum [61], or the Lightning state channels for Bitcoin [62]. |
| Data Storage: | Large data storage on-chain can be costly and bloat up the chain. Non-transactional data storage could be saved off-chain and linked to the DLT. This decision were to store data should be considered before selecting a DLT design option [2]. Some options for decentralized off-chain data storage already exist (e.g. IPFS [63], or bigchainDB [64]). |
| Interoperability: | Connection of the DLT with other parts of the technology stack [65] is very important for successful use cases. DLT either have no interoperability, explicit implemented tools to allow for interoperability or an implicit interoperability by connecting via smart contract to any API tool or interface [40]. This interoperability also involves connectivity to "oracles". Entries on the DLT do not verify the correctness of the data itself, it just promises that data cannot be altered. To securely bring data onto the ledger, so-called "oracles" are needed [66]. This can be human manual data input or data from sensors or third-party services. |
| Privacy: | Privacy is an important constraint. Businesses might not want to share data on public ledgers, or GDPR protections do not allow to make certain data publicly available. On-chain encryption can be |
| | an option (see section 3.1.1), but is sometimes also limited, since smart contracts cannot read and act upon encrypted information. Private permissionless systems might allow for more flexibility in this regard. |
| Smart Contracts: | If a use case relies on the use of smart contracts for automation or tokenization, the chosen DLT design should support computation on its application layer. In the future their might be also the possibility to add smart contracts retrospectively to a DLT that does currently not support smart contracts (e.g. as proposed in Wüst et al. [67]). |
| Cost Structure: | An existing DLT usually involves fees to pay for transactions. In contrast, a private network involves the initial investment costs of servers and the overhead costs in running the network, but often involves no transaction fees. Dependent on the chosen DLT design option, cost and capital structure might differ and affect the decision for a certain DLT design option. |

## 5. Analysis of Use Cases

Having identified the categories based on use-case clustering (Table 2), they are analyzed regarding suited DLT design options based on the framework introduced in Figure 2. The analysis was performed by the authors, following the rational of the framework by simulating and assuming possible use case constellations. Since the use cases are often described on a high level, sometimes multiple design options could be appropriate, dependent on the final constellation and relationships of the participants. In Figure 3, the nine combinations leading to a certain DLT design option after applying the framework to the analyzed use cases are pictured. Table 7 shows then the results for each use case after stage 1 and 2 of the framework. In the following sections, the analysis is discussed in more detail, going through the three stages of the introduced framework.

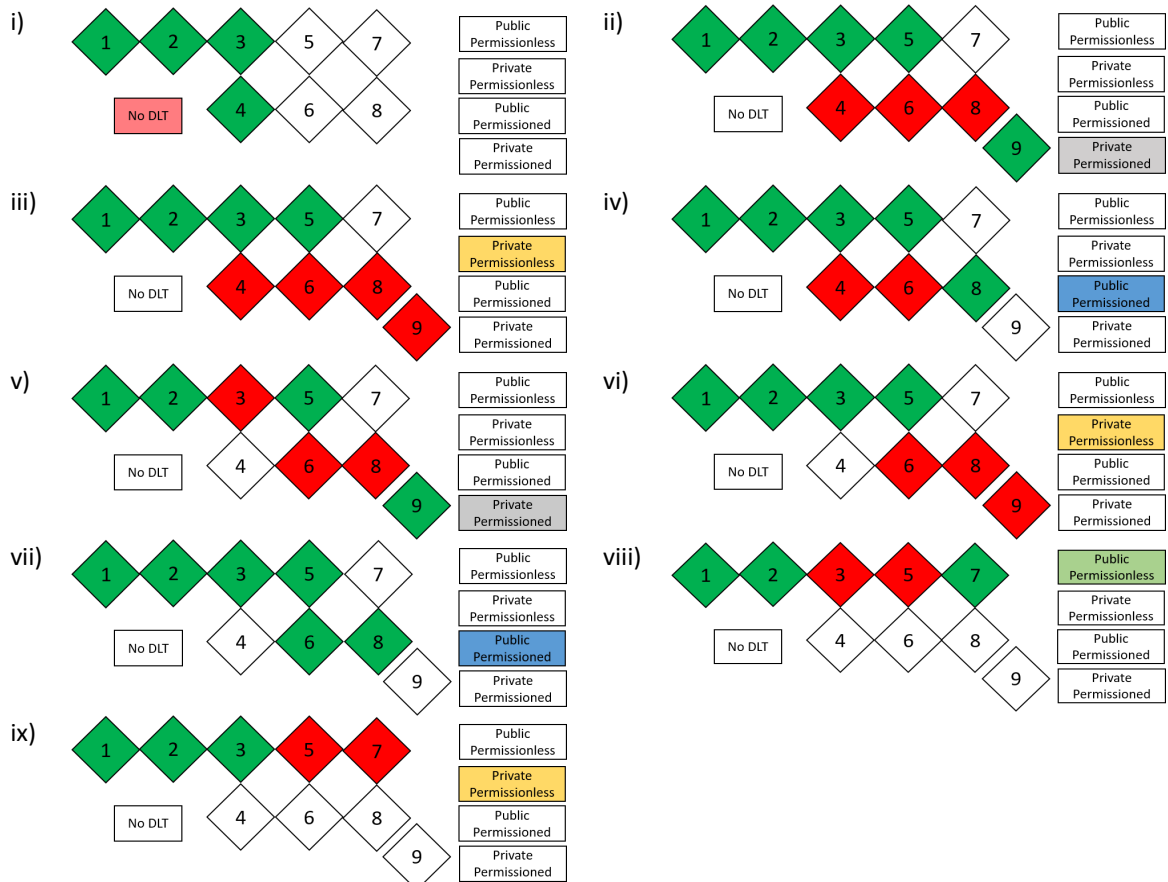

*Figure 3: Possible combinations for the analyzed use cases after stage 1 & 2 in the framework (Figure 2). Each rectangle stands for the respective question (1-9, see Figure 2), red for "no", and green for "yes".*

*Table 7: Results of applying stage 1 and 2 of the framework (Figure 2) to the identified use cases (Table 2), leading to possible combinations pictured in Figure 3.*

|   |   | No DLT | DLT (TTP possible) | | | DLT (TTP not possible) | | | | |
|---|---|---|---|---|---|---|---|---|---|---|
|   |   | i) | ii) | iii) | iv) | v) | vi) | vi) | vii) | ix) |
| 1 | **Internal Use for Administrative Processes** | | | | | | | | | |
| 1.1 | *Notarization and Synchronization of Documents* | X | X | X | | | | | | |
| 2 | **Transaction Automation between Stakeholders with Smart Contracts** | | | | | | | | | |
| 2.1 | *Triggering Payments* | X | X | X | X | | | | | |
| 2.2 | *Triggering Contract Deliverables* | X | X | X | X | | | | | |
| 2.3 | *Self-executing Contract Administration* | X | X | X | X | | | | | |
| 2.4 | *Automated Data/Information Sharing* | X | X | X | | | | | | |
| 2.5 | *Automated Code Compliance Checking* | X | X | X | X | | | | | |
| 3 | **Immutable Record of Transactions** | | | | | | | | | |
| 3.1 | *Timestamping of "Value" Transactions* | X | X | X | X | | | | | |
| 3.2 | *Record of Changes in digital models (BIM)* | X | X | X | X | | | | | |
| 3.3 | *Tracking of Supply Chain Logistics* | X | X | X | X | | | | | |
| 3.4 | *Tracking of Project Progress and Worked Hours* | X | X | X | X | | | | | |
| 3.5 | *Record of Maintenance and Operations Data* | X | X | X | X | | | | | |
| 3.6 | *Tracking of Health & Safety Incidents* | X | X | X | X | | | | | |
| 3.7 | *Verification of Installation Tasks* | X | X | X | X | | | | | |
| 3.8 | *Record/Notarization for Regulation and Compliance* | X | X | X | X | | | | | |
| 4 | **Immutable Record of Assets/Ownership** | | | | | | | | | |
| 4.1 | *Record of Ownership in BIM (IP-Rights)* | X | X | X | X | | | | | |
| 4.2 | *Record of Ownership for Physical Assets (e.g. Property)* | X | | | X | | | | | |
| 4.3 | *Managing Identities for Reputation (People, Contractors)* | X | | | X | | | | | |
| 4.4 | *Material & Product Passports (Provenance and Properties)* | X | | | X | | | | | |
| 5 | **Coins/Tokens as Payment or Incentive Scheme** | | | | | | | | | |
| 5.1 | *Payment in Cryptocurrencies* | | | | | | | | X | |
| 5.2 | *Shared Accounts & Insurances* | X | X | X | X | | | | | |
| 5.3 | *Incentives over the Whole Building-Lifecycle* | | | | | X | X | X | X | X |
| 6 | **Decentralized Applications (DApps)** | | | | | | | | | |
| 6.1 | *Decentralized Market Places for Products and Services* | | | | | | X | X | | |
| 6.2 | *Decentralized Common Data Environments (CDE) for Digital Models* | | | | | X | X | X | X | X |
| 7 | **Decentralized Autonomous Organizations (DAOs)** | | | | | | | | | |
| 7.1 | *Automated Building Maintenance Systems* | | | | | X | X | X | X | X |
16

*5.1. Stage 1 – Do you need DLT?*

All analyzed use cases need to store state and involve multiple writers (Figure 2, Question 1 & 2). Question 3 then asks whether an always online TTP can be used. This is definitely possible for many of the described use cases, especially for category 1 (*internal use for administrative purposes*), category 3 (*immutable record of transaction*), and category 4 (*immutable record of assets/ownership*). Also use case 5.2 (*shared account & insurances*) could make use of a third party operating this service. As soon as a TTP is possible, using DLT is not needed. However, there might be good reasons to still use DLT, such as reduced costs without a TPP or avoiding control by a TTP as an intermediary. Furthermore, the size and complexity of a solution might favor a decentralized network structure using DLT, e.g. in the case of supply chain tracking (Table 7, use case 3.3). Hence, if DLT seems desirable despite the possibility of having a TTP, question 4 (Figure 2, "Do you want to use a TTP?") can be answered with "no".

The framework then leads to the questions assessing the relationship of the involved participants, in particular whether they are known (Figure 2, Question 5) and whether their interests are aligned (Figure 2, Question 6). In the analyzed construction use cases, participants are generally known if a TTP is possible, so question 5 can always be answered with "yes". If also question 6 (interests not aligned) can be affirmed, the framework suggests to not use DLT. This acts as a fallback mechanism, even though not using a TTP was wanted (Figure 2, Question 4), since the drawbacks of using a DLT (in terms of performance, cost, or other) is most likely not justified. Since the exact relationship of participants in the analyzed use cases was in general not described and at least a non-alignment of interests was possible, this combination was not considered in Figure 3 and Table 7.

For some use cases a TTP is not possible, directly leading to the evaluation of relationships between the participants (Figure 2, Question 5 & 6). This applied to use cases that rely on some functionality of a DLT, such as the payment in cryptocurrencies (Table 7, use case 5.1 & 5.3) or the decentralized characteristics of the solution (Table 7, categories 6 & 7). These are also use cases that do not already exist in construction, but would be knew solutions enabled through DLT.

*5.2. Stage 2 – Which DLT design option?*

If a DLT is a suited solution after stage 1, the final DLT design option depends on whether the starting point for the assessment is that participants are unknown (Figure 2, Question 5) or interests are not aligned (Figure 2, Question 6). If participants are known but interests not aligned (mostly the case if a TTP is possible), three options ii), iii), and iv) in Figure 3 need to be considered. If participants are unknown (mostly the case if a TTP is not possible), both question 5 & 6 appear as starting points, depending on the specific relationship of participants. Often, the use cases were not described in enough detail, so both options had to be considered, leading to five possible combinations v) to ix) in Figure 3.

As a next step, question 7 & 8 (Figure 2) filter use cases where public verifiability is required or wanted. E.g. for use case 2.4 (*Automated Data/Information Sharing*), most likely no public verifiability is wanted, since the documents can contain sensitive information. A similar situation is use case 1.1 (*Notarization and Synchronization of Documents*), where documents are only shared internally. In contrast, use case category 4 (*Immutable Record of Assets/Ownership*) most likely requires public verifiability to ensure trust and transparency to outside parties. For use case 4.1 (*Record of Ownership in BIM (IP-Rights)*), it could be both depending on the needs of the involved parties, since ownerships in a BIM model could also be managed internally. Similarly, for all the other use cases were a TTP is possible, public verifiability could be either wanted or not depending on the details of the use case. Looking at use cases were it was assessed that a TTP is not possible, it is clear that use case 5.1 (*Payment in Cryptocurrencies*) and 6.1 (*Decentralized Market Places for Products and Services*) both need public verifiability to be trustworthy. For all other use cases, it cannot be finally assessed based on the provided information whether public verifiability is needed or not.

The last question in stage 2 assesses if control of functionality on the protocol level is required (Figure 2, Question 9) in case of a private DLT option. This highly depends on the parties involved in a use case and their preferences in how to set up a private DLT. Therefore, always both combinations (Figure 3, ii/iii & v/vi) were marked as possible options (Table 7).

*5.3. Stage 3 – Constraints?*

In stage 3 additional constraints relevant to the final DLT design options should be discussed (see Figure 2). Since this is an assessment based on the final relationship of involved parties in the specific use case and the proposed DLT design option resulting from stage 2, it was not possible to facilitate specific discussions without further specification of the use cases. An exemplary discussion around some possible constraints is provided to clarify the procedure for use case category 2. A detailed assessment of constraints would need to be conducted for each final use case. If the proposed DLT design option after stage 2 cannot be realized, or other constraints are more important, another DLT option or no DLT might be chosen.

Example: Category 2 (Table 7, *Transaction Automation between Stakeholders with Smart Contracts*) needs to consider constraints related to smart contracts. First, the chosen DLT needs to support *smart contracts* on the application layer. If the purpose of a smart contract is to act on external state information in the ledger, a publicly verifiable system that replicates data on all nodes is needed (*interoperability*). *Throughput* might be an issue if many smart contract interactions are needed. Private DLT generally provide better performance. Alternatives would be to use $2^{nd}$ layer solutions for public DLT. Regarding *privacy*, on-chain encryption in public systems would in most cases not allow a smart contract to execute logic based on that data. Private DLT would still allow for privacy, but mostly come with less security. And since all DLT design options are possible, preferences regarding the different *cost structures* should be considered.

## 6. Summary and Discussion

This section was structured according to the three main contributions of this paper: 1) the categorization of use cases in construction, 2) the introduced framework to choose a DLT design option for a specific use case, and 3) the analysis of the reviewed use cases with the proposed framework.

*6.1. Use Case Categorization*

*6.1.1. Contribution*

DLT use cases in construction were summarized from state-of-the art literature, extending the work of Hunhevicz and Hall [27]. A more detailed assessment with the new framework allowed the identification of an additional use case category and some relocations of use cases to another category. The reviewed use cases show the broad potential application field of DLT use cases in construction, of which many promise improvements regarding transparency and process optimizations through automation and disintermediation. While not identifying many new use cases compared to reviews in past literature (e.g. in Li et al. [9]), the categorization according to specific value propositions of DLT can lead to a more structured thinking and better overview of the commonalities and differences between construction DLT use cases. This can be particularly helpful in the decision process when trying to implement the use case with the best-suited DLT design option.

*6.1.2. Limitations*

For the purpose of this paper, even though trying to include all relevant literature, no systematic literature review was conducted. Therefore, there is no claim in being complete with the identified use cases. Moreover, because of the early state of research, it is expected that the use case categorization is subject to change while the use cases and technology evolve. If needed, the use cases and categories should be revised or extended.

*6.1.3. Future research*

Considering the early stage of DLT research in construction and its manifold applications, there is potential to identify additional and innovative use cases of DLT in construction. The authors expect that more use cases will be introduced as a refinement or combination of different use cases. Especially the categories 6 (Table 2, *Dapp*) and 7 (Table 2, *DAO*) will likely grow in importance as a combination and extension of use cases. E.g. Li et al. [9] mention single shared access BIM models as a combination of use case 3.2 (Table 2, *record of changes in BIM*), 4.1 (Table 2, *record of ownership in BIM*), and 6.3 (Table 2, *Decentralized data storage*). Having said that, there seems to be a tendency to apply DLT to existing processes in construction, which raises the question about the actual benefits in comparison. There is need to move beyond the theorization of use cases towards prototypes and case studies to further advance the research in this field. Either to quantitatively compare the existing processes with and without an implementation of DLT, or to showcase and assess the benefit and change to construction processes through innovative and new use cases enabled by DLT.

## 6.2. Decision Framework for DLT Design Options

### 6.2.1. Contribution

A framework was introduced to link the use cases to DLT design options. Eight existing frameworks were reviewed and cross-compared (see Table 5). This allowed to supplement the various frameworks with aspects not considered previously, while prioritizing points that were considered more often. The final logic of the framework is based on what fundamental properties of a DLT design option are required for a given use case, optimizing the performance of the chosen DLT design option (stage 1 & 2). Since the different DLT design options always compromise one or the other aspect, it is important to consider constraints in stage 3. This allows to readjust the technical solution to factors that might be limiting or of higher importance for certain use cases. In contrast to the reviewed frameworks that also consider some technical constraints (e.g. Rangaswami et al. [58] and Li et. al [9]), the proposed framework determines first whether DLT would be suited based on the fundamental properties and only then assesses various constraints. The authors expect that this will lead to longer validity of the framework, since the fundamental properties of DLT are not expected to change as fast as technical constraints. Finally, in addition to the underlying framework of Wüst and Gervais [60], the authors included also the emerging design option of *private permissionless* DLT to be complete in the currently available DLT design options.

### 6.2.2. Limitations

The proposed framework is based on the reviewed frameworks in Table 5, highlighting the theoretical connection between the trust relationships of participants in a use cases and the varying fundamental properties of DLT design options. This theoretical connection should be verified with future practical implementation. Furthermore, while stage 1 and 2 guide the reader through the different aspects without much knowledge about DLT, a potential limitation is that stage 3 requires in-depth technical knowledge of the user to assess the different constraints.

### 6.2.3. Future research

Future research should examine how to create more extensive frameworks to decide for a certain DLT design option. One potential starting point could be a structured decision tree for stage 3 (similar to stage 1 and 2). Furthermore, as more combined DLT use cases emerge (e.g. within one construction project), the question arises how to deal with the potentially different technical prerequisites between them. For that, emerging hybrid solutions combining different DLT design options could be considered in the framework. Having said that, some hybrid solutions might be categorized in the private permissionless DLT design option, such as Ark [68] or LTO network [69], and might therefore already be implicitly considered. Moreover, the emerging complementary technology stack (see e.g. Web3 Hub [65]) together with existing software solutions used in construction could be included when searching for the best possible technical solutions for a use case. Finally, once a DLT design option was chosen with the framework, a product in the market needs to be selected for implementation. Future research should list and look at these products and map them to the different DLT design options, highlighting also specific constraints.

## 6.3. Use Case Analysis

### 6.3.1. Contribution

The introduced framework was used to classify DLT design options of proposed use cases in construction. The main contribution here is that the assessment can hint whether or not DLT would be a good solution for use cases in construction based on the need for a trusted solution, and if true, which specific DLT design option should be chosen.

Regarding whether DLT would be a good solution, the analysis of the use cases with the framework indicate at least that the fundamental properties provided by DLT could be beneficial for the described use cases. Having said that, for many of the described use case a trusted third party (TTP) would be possible to achieve the same result. This means a DLT would not necessarily be needed. In general, this was found to be true if DLT should be applied to existing processes. This does not mean there are no benefits by using DLT in these cases. It is then up to the more detailed assessment whether the savings from not having a TTP justify the cost of having a DLT. Only few of the proposed use cases actually require the use of DLT. Often, they are described even more high level than the use of DLT in existing solutions. Overall, despite a theoretical alignment of DLT fundamental properties and use case requirements, it is currently not possible to assess if and to what extent DLT use cases benefit construction. It seems that a better answer to "Do you need a blockchain (or another type of DLT) in construction?" can only be given once prototypes have been built and the benefits have been validated through case studies.

Regarding the best-suited DLT design options, the framework results in more than one possible option for most of the considered use cases. This is likely due to the fact that use cases are not described in enough detail. In a specific implementation of DLT for construction, the best-suited DLT design option will be dependent on the final constellation of participants. Having said that, there is some consistency of possible DLT design options recognizable within the categories.

*6.3.2. Limitations*

Even though the performed use case analysis can help to understand potential DLT design options for individual use cases, the picture is somewhat diluted and needs further refinement. This is mostly due to the fact that the participants' trust relationship was mostly hard to assess with the provided use case descriptions. Hunhevicz and Hall [27] expected that the different use case categories will have an increasing need towards higher fundamental properties with decreasing level of trust. Looking at the performed classification (Table 7), this relationship could not be clearly recognized. Multiple DLT design options are possible for most use cases without better specifications between the participant's trust relationships. Finally, the analysis was performed by the authors with the best of knowledge about the use case constellations and should be verified by construction industry experts and DLT domain experts.

*6.3.3. Future research*

As mentioned in the limitations, most use cases do not describe the exact relationship of participants, which would be important to assess the best-suited DLT design option. Therefore, more in-depth analysis of use cases and the relationships of the participants is needed in future research for a more insightful analysis and classification of suited DLT design options. Moreover, there might be barriers for future use case implementation related to other socio-technical challenges that should be also carefully studied. A starting point for this could be the framework of implementation challenges by Li et al. [9] in four dimensions (technical, process, social, policy). Finally, the use case analysis is based on current processes in construction. Having a DLT solution in place could potentially change processes and the relationship of participating parties, which would lead to a different assessment using the framework (e.g. allowing unknown parties to participate in a construction process). Future research could try to incorporate and analyze these relationships.

## 7. Conclusion

This paper structured and assessed use cases in construction for blockchain and other types of distributed ledger technology (DLT) regarding their actual need for such a technical solution. For this, an overarching decision framework based on previous work was introduced to link use cases to four DLT design options according to the needed fundamental properties of a use case.

Indeed, many of the analyzed construction use cases could potentially profit from using DLT. However, most of the use cases applied DLT to existing processes, where a DLT is not necessarily required. In these cases, further investigation is needed whether the added value of having a DLT justifies its application. Only few proposals used DLT as a tool to enable innovative use cases that cannot be realized without DLT. For a better perspective on whether DLT can be overall beneficial for the construction industry, more in-depth analysis of the use cases is needed regarding their added value and socio-economic impacts, best trough prototypes and case studies. For that the different possible DLT design options should be considered, since the proposed use cases in construction seem to vary considerably in the constellation of trust relationship among participants. However, this was found to be challenging, since most use cases do not describe the exact relationship of participants, which would be important to assess the best-suited DLT design option. More in-depth analysis of use cases and the relationships of the participants is needed for a final assessment.

Nevertheless, the at least partial alignment of construction use cases with fundamental properties of DLT should encourage researchers and practitioners to further explore the topic. For that the use case clustering together with the introduced framework is expected to act as a valuable tool to think more interconnected between use cases in construction and DLT design options to advance the research in this field.

**Funding**

This research did not receive any specific grant from funding agencies in the public, commercial, or not-for-profit sectors.

**References**

[1] S. Nakamoto, Bitcoin: A Peer-to-Peer Electronic Cash System, Www.Bitcoin.Org. (2008). https://doi.org/10.1007/s10838-008-9062-0.

[2] X. Xu, I. Weber, M. Staples, L. Zhu, J. Bosch, L. Bass, C. Pautasso, P. Rimba, A Taxonomy of Blockchain-Based Systems for Architecture Design, in: 2017 IEEE Int. Conf. Softw. Archit., IEEE, 2017: pp. 243–252. https://doi.org/10.1109/ICSA.2017.33.

[3] G. Hileman, M. Rauchs, 2017 Global Blockchain Benchmarking Study, SSRN Electron. J. (2017). https://doi.org/10.2139/ssrn.3040224.

[4] N. El Ioini, C. Pahl, A Review of Distributed Ledger Technologies, in: Springer, Cham, 2018: pp. 277–288. https://doi.org/10.1007/978-3-030-02671-4_16.

[5] V. Buterin, Ethereum: A Next-Generation Smart Contract and Decentralized Application Platform, White Pap. (2014). https://github.com/ethereum/wiki/wiki/White-Paper (accessed August 27, 2018).

[6] W. Viryasitavat, L. Da Xu, Z. Bi, A. Sapsomboon, Blockchain-based business process management (BPM) framework for service composition in industry 4.0, J. Intell. Manuf. (2018) 1–12. https://doi.org/10.1007/s10845-018-1422-y.

[7] D. Tapscott, A. Tapscott, How Blockchain Will Change Organizations, 58 (2016) 10–13. https://sloanreview.mit.edu/article/how-blockchain-will-change-organizations/ (accessed July 30, 2018).

[8] W. Nowiński, M. Kozma, How Can Blockchain Technology Disrupt the Existing Business Models?, Entrep. Bus. Econ. Rev. 5 (2017) 173–188. https://doi.org/10.15678/EBER.2017.050309.

[9] J. Li, D. Greenwood, M. Kassem, Blockchain in the built environment and construction industry: A systematic review, conceptual models and practical use cases, Autom. Constr. 102 (2019) 288–307. https://doi.org/10.1016/J.AUTCON.2019.02.005.

[10] B. Penzes, Blockchain technology: could it revolutionise construction?, Institution of Civil Engineers, 2018. https://www.ice.org.uk/news-and-insight/the-civil-engineer/december-2018/can-blockchain-transform-construction (accessed December 30, 2018).

[11] A. Dubois, L.-E. Gadde, The construction industry as a loosely coupled system: implications for productivity and innovation, Constr. Manag. Econ. 20 (2002) 621–631. https://doi.org/10.1080/01446190210163543.

[12] R. Zolin, P.J. Hinds, R. Fruchter, R.E. Levitt, Interpersonal trust in cross-functional, geographically distributed work: A longitudinal study, Inf. Organ. 14 (2004) 1–26. https://doi.org/10.1016/j.infoandorg.2003.09.002.

[13] D.M. Hall, The Joint Impact of Supply Chain Integration Practices on Construction Schedule Performance for California Healthcare Projects, in: Constr. Res. Congr. 2018 Infrastruct. Facil. Manag. - Sel. Pap. from Constr. Res. Congr. 2018, 2018. https://doi.org/10.1061/9780784481295.019.

[14] P. Pishdad-Bozorgi, Y.J. Beliveau, Symbiotic Relationships between Integrated Project Delivery (IPD) and Trust, Int. J. Constr. Educ. Res. 12 (2016) 179–192.

https://doi.org/10.1080/15578771.2015.1118170.

[15] Tavistock Institute of Human Relations, Interdependence and Uncertainty: A Study of the Building Industry. Digest of a Report from the Tavistock Institute to the Building Industry Communication Research Project, Tavistock Pubs., London, UK, 1966.

[16] D. Hall, A. Algiers, T. Lehtinen, R.E. Levitt, C. Li, P. Padachuri, The Role of Integrated Project Delivery Elements in Adoption of Integral Innovations, in: P. Chan, R. Leicht (Eds.), Eng. Proj. Organ. Conf. 2014 Proc., Engineering Project Organization Society, Devil's Thumb Ranch, Colorado, 2014: pp. 1–20.

[17] M. Mathews, D. Robles, B. Bowe, BIM+Blockchain: A Solution to the Trust Problem in Collaboration?, CITA BIM Gather. (2017). https://arrow.dit.ie/bescharcon/26 (accessed July 23, 2018).

[18] C. Kinnaird, M. Geipel, Blockchain Technology: How the Inventions Behind Bitcoin are Enabling a Network of Trust for the Built Environment, Arup, 2017. https://www.arup.com/perspectives/publications/research/section/blockchain-technology (accessed July 24, 2018).

[19] Z. Ye, M. Yin, L. Tang, H. Jiang, Cup-of-Water theory : A review on the interaction of BIM , IoT and blockchain during the whole building lifecycle, (2018).

[20] E. Papadonikolaki, H. Wamelink, Inter- and intra-organizational conditions for supply chain integration with BIM, Build. Res. Inf. 45 (2017) 649–664. https://doi.org/10.1080/09613218.2017.1301718.

[21] R. Miettinen, S. Paavola, Beyond the BIM utopia: Approaches to the development and implementation of building information modeling, Autom. Constr. 43 (2014) 84–91. https://doi.org/10.1016/J.AUTCON.2014.03.009.

[22] A. Ghaffarianhoseini, J. Tookey, A. Ghaffarianhoseini, N. Naismith, S. Azhar, O. Efimova, K. Raahemifar, Building Information Modelling (BIM) uptake: Clear benefits, understanding its implementation, risks and challenges, Renew. Sustain. Energy Rev. 75 (2017) 1046–1053. https://doi.org/10.1016/J.RSER.2016.11.083.

[23] D.M. Hall, W.R. Scott, Early Stages in the Institutionalization of Integrated Project Delivery, Proj. Manag. J. 50 (2019) 875697281881991. https://doi.org/10.1177/8756972818819915.

[24] E. Papadonikolaki, Loosely Coupled Systems of Innovation: Aligning BIM Adoption with Implementation in Dutch Construction, J. Manag. Eng. 34 (2018) 05018009. https://doi.org/10.1061/(ASCE)ME.1943-5479.0000644.

[25] E. Fleisch, What is the Internet of Things ? - An Economic Perspective, Auto-ID Labs White Pap. WP-BIZAPP-053. 5 (2010) 1–27. https://doi.org/10.1109/MCOM.2013.6476854.

[26] J. De La Pena, E. Papadonikolaki, From relational to technological trust: How do the Internet of Things and Blockchain technology fit in?, in: 2019: pp. 415–424. https://doi.org/10.35490/EC3.2019.153.

[27] J.J. Hunhevicz, D.M. Hall, Managing mistrust in construction using DLT: a review of use-case categories for technical decisions, in: 2019 EC3 Conf. Greece, 2019: pp. 100–109. https://doi.org/10.35490/EC3.2019.171.

[28] I. Belle, The architecture, engineering and construction industry and blockchain technology, JI, G. & TONG, Z. (eds.) Digital Culture 数码文化 Proceedings of 2017 National Conference on Digital Technologies in Architectural Education, Nanjing: China Architecture Industry Publishers., 2017.

[29] A. Heiskanen, The technology of trust: How the Internet of Things and blockchain could usher in a new era of construction productivity, Constr. Res. Innov. 8 (2017) 66–70. https://doi.org/10.1080/20450249.2017.1337349.

[30] D. Kifokeris, C. Koch, Blockchain in construction logistics: state-of-art, constructability, and the advent of a new digital business model in Sweden, in: 2019: pp. 332–340. https://doi.org/10.35490/EC3.2019.163.

[31] J. Li, M. Kassem, A. Ciribini, M. Bolpagni, A Proposed Approach Integrating DLT, BIM, IoT and Smart Contracts: Demonstration Using a Simulated Installation Task, in: Int. Conf. Smart Infrastruct. Constr. 2019, 2019. https://doi.org/10.1680/ICSIC.64669.275.

[32] H. Luo, M. Das, J. Wang, J.C.P. Cheng, H. Kong, Construction Payment Automation through Smart Contract-based Blockchain Framework, in: ISARC. Banff, Canada, 2019: pp. 1–7.

[33] J. Mason, Intelligent Contracts and the Construction Industry, J. Leg. Aff. Disput. Resolut. Eng. Constr. 9 (2017) 04517012. https://doi.org/10.1061/(ASCE)LA.1943-4170.0000233.

[34] N.O. Nawari, S. Ravindran, Blockchain and Building Information Modeling (BIM): Review and Applications in Post-Disaster Recovery, Buildings. 9 (2019) 149. https://doi.org/10.3390/buildings9060149.

[35] A. O'Reilly, M. Mathews, Incentivising Multidisciplinary Teams with New Methods of Procurement using BIM + Blockchain, Conf. Pap. (2019). https://arrow.dit.ie/bescharcon/31 (accessed June 11, 2019).

[36] Ž. Turk, R. Klinc, Potentials of Blockchain Technology for Construction Management, Procedia Eng. 196 (2017) 638–645. https://doi.org/10.1016/J.PROENG.2017.08.052.

[37] J. Wang, P. Wu, X. Wang, W. Shou, The outlook of blockchain technology for construction engineering management, Front. Eng. Manag. 4 (2017) 67. https://doi.org/10.15302/J-FEM-2017006.

[38] R. Wattenhofer, Distributed Ledger Technology: The Science of the Blockchain (2nd ed.), CreateSpace Independent Publishing Platform, USA, 2017. https://dl.acm.org/citation.cfm?id=3153775 (accessed November 16, 2018).

[39] M.C. Ballandies, M.M. Dapp, E. Pournaras, Decrypting Distributed Ledger Design - Taxonomy, Classification and Blockchain Community Evaluation, (2018). http://arxiv.org/abs/1811.03419 (accessed November 19, 2018).

[40] P. Tasca, C.J. Tessone, Taxonomy of Blockchain Technologies. Principles of Identification and Classification, (2017). http://arxiv.org/abs/1708.04872 (accessed November 13, 2018).

[41] V. Shermin, Disrupting governance with blockchains and smart contracts, Strateg. Chang. 26 (2017) 499–509. https://doi.org/10.1002/jsc.2150.

[42] S. Popov, IOTA whitepaper v1.4.3, New Yorker. 81 (2018) 1–28. https://assets.ctfassets.net/r1dr6vzfxhev/2t4uxvsIqk0EUau6g2sw0g/45eae33637ca92f85dd9f4a3a218e1ec/iota1_4_3.pdf.

[43] S.A. Back, M. Corallo, L. Dashjr, M. Friedenbach, G. Maxwell, A. Miller, A. Poelstra, J. Timón, Enabling Blockchain Innovations with Pegged Sidechains, (2014). https://www.semanticscholar.org/paper/Enabling-Blockchain-Innovations-with-Pegged-Back-Corallo/1b23cd2050d5000c05e1da3c9997b308ad5b7903 (accessed November 23, 2018).

[44] M. Zamani, M. Movahedi, M. Raykova, RapidChain, in: Proc. 2018 ACM SIGSAC Conf. Comput. Commun. Secur. - CCS '18, ACM Press, New York, New York, USA, 2018: pp. 931–948. https://doi.org/10.1145/3243734.3243853.

[45] J. Yocom-Piatt, Surveying the Privacy Landscape, (2019). https://blog.decred.org/2019/08/21/Surveying-the-Privacy-Landscape/ (accessed November 11, 2019).

[46] N. Szabo, Smart Contracts: Building Blocks for Digital Markets, (1996). http://www.fon.hum.uva.nl/rob/Courses/InformationInSpeech/CDROM/Literature/LOTwinterschool2006/szabo.best.vwh.net/smart_contracts_2.html (accessed November 23, 2018).

[47] Token Alliance, Understanding Digital Tokens: Market Overviews and Proposed Guidelines for Policymakers and Practitioners, Chamber of Digital Commerce, 2018. https://digitalchamber.org/token-alliance-whitepaper/ (accessed November 14, 2018).

[48] R.G. Brown, J. Carlyle, I. Grigg, M. Hearn, Corda : An Introduction, (2016) 1–15.

[49] E. Harris-braun, N. Luck, A. Brock, Holochain, (2018) 1–14.

[50] A. Gervais, G.O. Karame, K. Wüst, V. Glykantzis, H. Ritzdorf, S. Capkun, On the Security and Performance of Proof of Work Blockchains, in: Proc. 2016 ACM SIGSAC Conf. Comput. Commun. Secur. - CCS'16, ACM Press, New York, New York, USA, 2016: pp. 3–16. https://doi.org/10.1145/2976749.2978341.

[51] B.S. Voshmgir, M. Zargham, Foundations of Cryptoeconomic Systems, (2019) 1–18.

[52] M. Castro, B. Liskov, Practical Byzantine Fault Tolerance, Proc. Symp. Oper. Syst. Des. Implement. (1999) 1–14. https://doi.org/10.1145/571637.571640.

[53] G. Miscione, T. Goerke, S. Klein, G. Schwabe, R. Ziolkowski, Hanseatic Governance: Understanding Blockchain as Organizational Technology, ICIS 2019 Proc. (2019). https://aisel.aisnet.org/icis2019/blockchain_fintech/blockchain_fintech/3 (accessed November 14, 2019).

[54] P. Maymounkov, D. Mazières, Kademlia: A Peer-to-Peer Information System Based on the XOR Metric, in: Springer, Berlin, Heidelberg, 2002: pp. 53–65. https://doi.org/10.1007/3-540-45748-8_5.

[55] A. Daniels, The rise of private permissionless blockchains — part 1, (2018). https://medium.com/ltonetwork/the-rise-of-private-permissionless-blockchains-part-1-4c39bea2e2be (accessed August 29, 2019).

[56] M.E. Peck, Do you need a blockchain? This chart will tell you if the technology can solve your problem, IEEE Spectr. 54 (2017) 38–60. https://doi.org/10.1109/MSPEC.2017.8048838.

[57] B. Suichies, Why Blockchain must die in 2016, (2015). https://medium.com/block-chain/why-blockchain-must-die-in-2016-e992774c03b4 (accessed August 29, 2019).

[58] J. Rangaswami, S. Warren, C. Mulligan, J. Zhu Scott, Blockchain Beyond the Hype A Practical Framework for Business Leaders, White Pap. World Econ. Forum 2018. (2018). http://www3.weforum.org/docs/48423_Whether_Blockchain_WP.pdf%0Ahttp://www3.weforum.org/docs/48423_Whether_Blockchain_WP.pdf%0Ahttps://www.weforum.org/agenda/2018/04/questions-blockchain-toolkit-right-for-business.

[59] F. Wessling, E. Christopher, M. Hesenius, V. Gruhn, How Much Blockchain Do You Need? Towards a Concept for Building Hybrid DApp Architectures Florian, in: 2018 IEEE/ACM 1st Int. Work. Emerg. Trends Softw. Eng. Blockchain, 2018: pp. 44–47. https://ieeexplore.ieee.org/abstract/document/8445058 (accessed August 22, 2019).

[60] K. Wust, A. Gervais, Do you Need a Blockchain?, in: 2018 Crypto Val. Conf. Blockchain Technol., IEEE, 2018: pp. 45–54. https://doi.org/10.1109/CVCBT.2018.00011.

[61] J. Poon, V. Buterin, Plasma: Scalable Autonomous Smart Contracts, Whitepaper. (2017) 1–47. https://plasma.io/plasma.pdf%0Ahttps://plasma.io/%0Ahttps://plasma.io/plasma.pdf.

[62] J. Poon, T. Dryja, The Bitcoin Lightning Network: Scalable Off-Chain Instant Payments, (2016). https://lightning.network/lightning-network-paper.pdf.

[63] J. Benet, IPFS - Content Addressed, Versioned, P2P File System, (2014). http://arxiv.org/abs/1407.3561 (accessed November 11, 2019).

[64] T. Mcconaghy, R. Marques, A. Müller, D. De Jonghe, T. Mcconaghy, G. Mcmullen, R. Henderson, S. Bellemare, A. Granzotto, BigchainDB: A Scalable Blockchain Database (DRAFT), BigchainDB. (2016) 1–65.

[65] Web3 Hub, Tech Stack Overview, (2019).

http://wiki.web3.foundation/en/latest/tech_stack/tech_stack_overview/ (accessed November 11, 2019).

[66] X. Xu, C. Pautasso, L. Zhu, V. Gramoli, A. Ponomarev, A.B. Tran, S. Chen, The Blockchain as a Software Connector, in: 2016 13th Work. IEEE/IFIP Conf. Softw. Archit., IEEE, 2016: pp. 182–191. https://doi.org/10.1109/WICSA.2016.21.

[67] K. Wüst, L. Diana, K. Kostiainen, G. Karame, S. Matetic, S. Capkun, Bitcontracts: Adding Expressive Smart Contracts to Legacy Cryptocurrencies, (2019) 1–14. https://eprint.iacr.org/2019/857.pdf.

[68] Ark, Ark Ecosystem Whitepaper, (2019). https://ark.io/Whitepaper.pdf.

[69] LTO.network, Blockchain for Decentralized Workflows, (2019). https://ltonetwork.com/documents/LTO Network - Technical Paper.pdf.